\documentclass[preprint,12pt]{elsarticle}
\biboptions{sort&compress}
\usepackage{times} 
\usepackage{amsmath} 
\usepackage{amssymb}  
\usepackage{psfrag}
\usepackage{latexsym,graphicx}
\newtheorem{theorem}{Theorem}

\newtheorem{definition}{Definition}

\newtheorem{remark}{Remark}
\journal{Systems \& Control Letters}

\begin{document}

\begin{frontmatter}

\title{Low Frequency Approximation for a class of Linear  Quantum Systems using Cascade Cavity Realization\tnoteref{ARC}}
 \tnotetext[ARC]{This work was supported by the Australian Research Council and Air Force Office of Scientific Research (AFOSR). This material is based on research sponsored by the Air Force Research Laboratory, under agreement number FA2386-09-1-4089.  The U.S. Government is authorized to reproduce and distribute reprints for Governmental purposes notwithstanding any copyright notation thereon.
The views and conclusions contained herein are those of the authors and should not be interpreted as necessarily representing the official policies or endorsements, either expressed or implied, of the Air Force Research Laboratory or the U.S. Government.}
\author{Ian R.~Petersen}

\address{School of Engineering and Information Technology, 
University of New South Wales at the Australian Defence Force Academy,
Canberra ACT 2600, Australia.}
\ead{i.r.petersen@gmail.com}

\begin{abstract}
This paper presents a method for approximating a class of complex
transfer function matrices corresponding to physically
realizable complex linear quantum systems. 
The class of  linear quantum systems under consideration
includes interconnections of passive optical components such as
cavities, beam-splitters, phase-shifters and interferometers. This approximation method builds on a previous result for  cascade realization  and gives good approximations at low frequencies. 
\end{abstract}


\end{frontmatter}

\section{Introduction} \label{sec:intro}
In recent years, there has been considerable interest in the  modeling  and feedback
control of
linear quantum systems; e.g., see
\cite{YK03A,YK03B,YAM06,JNP1,NJP1,GGY08,MaP3,MaP4,MAB08,YNJP1,PET08Aa,PET08A,ShP5a,NJD09,GJ09,GJN10,WM10}.
Such linear quantum  
systems commonly arise in the area of quantum optics; e.g., see
\cite{WM94,BR04}. The feedback control of quantum optical systems has
applications in areas such as quantum communications, quantum 
teleportation, and gravity wave detection. In particular, the papers
\cite{MaP3,MaP4,PET08Aa,PET08A} have been concerned with a class of linear quantum
systems in which the system can be defined in terms of a set of linear
complex quantum stochastic differential equations (QSDEs) defined
purely in terms of annihilation operators. Such linear complex quantum
systems correspond to optical systems made up of passive optical
components  such as optical cavities, beam-splitters, and phase
shifters. This paper is concerned with the approximation of systems in this
class of linear complex quantum systems. The method proposed
in this paper builds on the result of \cite{PET08Aa,PET08A} which gives a
method for physically realizing a given  complex transfer function
matrix corresponding to a linear quantum system in the class
considered in \cite{MaP3,MaP4}. 

In the approximation of  linear quantum systems, it is important that
the approximate system which is obtained is physically realizable. The
issue of physical realizability for 
linear quantum systems was considered in the papers
\cite{JNP1,NJP1,MaP3,MaP4,ShP5a}. This notion relates to whether a given
QSDE model represents a physical quantum system which obeys the laws
of quantum mechanics. In applying applying approximation methods to
obtain approximate models of quantum systems, it is important that the
approximate model obtained is a physically realizable quantum
system so that it exhibits the features inherent to quantum mechanics such as the Heisenberg uncertainty principle. 

The approximation method proposed in this paper follows directly from
the physical relation algorithm proposed in \cite{PET08Aa,PET08A}. The
physical realizability of the approximate system follows directly from
the fact that the algorithm proposed in \cite{PET08Aa,PET08A} leads to a
physical realization in terms of a cascade connection of optical
cavities. For this approximation method,  we present some
bounds and approximate bounds on the approximation error as a function
of frequency. 

One application of the approximation method proposed in this paper
is in modelling of linear quantum systems where it is
desired to construct a simpler, but still physically realizable model
of a complex quantum linear system in such a way that a frequency
dependent bound on the approximation error is obtained at low
frequencies. Another application of the approximation method occurs in
the case of coherent quantum feedback control systems when both the plant and
controller are linear quantum systems; see
\cite{YK03A,YK03B,JNP1,NJP1a,MaP4}. In this case, it is desired to
construct a simpler coherent quantum controller which is still
physically realizable.


\section{A Class of Linear Complex Quantum Systems} \label{sec:systems}
We  consider a class of  linear quantum  systems described in terms of the
 annihilation operator by the  quantum stochastic differential
equations (QSDEs):
\begin{eqnarray}\nonumber
\label{sys0}
 da(t) &=&  F a(t)dt +    G du(t);   \nonumber \\
 dy(t) &=&  \bar H a(t)dt + J du(t)
\end{eqnarray}
where $F \in \mathbb{C}^{n \times n}$, $ G \in \mathbb{C}^{n
\times m}$, $\bar H \in \mathbb{C}^{m \times n}$ and
$J \in \mathbb{C}^{m \times m}$; e.g., see
\cite{MaP3,MaP4,JNP1,GZ00,BR04,WM10}. Here $ a(t) = \left[ {a_1 (t) 
\cdots a_n (t)} \right]^T$ is a vector of (linear combinations of) annihilation operators.
The vector $u(t)$ represents the input signals and is assumed to admit the
decomposition:
$
 du(t) = \beta _{u}(t)dt + d\tilde u(t)
$
where $\tilde u(t)$ is the noise part of $u(t)$ and $\beta_{u}(t)$
is an adapted process (see \cite{BHJ07},
\cite{PAR92} and \cite{HP84}).
The noise $\tilde u(t)$ is a vector of quantum noises.  The noise
processes can be represented as operators on an appropriate Fock 
space (for more details see \cite{VPB92},
\cite{PAR92}).
The process $\beta_{u}(t)$ represents variables of other systems
which may be passed to the system (\ref{sys0}) via an interaction. More
details on this class of  systems can be found in  \cite{MaP3}, \cite{JNP1}.

\begin{definition} (See \cite{MaP3,GGY08}.) 
\label{phys_real}
A  linear quantum 
  system of the form (\ref{sys0}) is 
  said to be {\em physically realizable} if there exists a commutation
  matrix $\Theta=\Theta^\dagger > 0$, a coupling matrix $\Lambda$, 
  a Hamiltonian matrix $M=M^\dagger$, and a scattering matrix $S$ such that
\begin{eqnarray}
\label{harmonic}
F  &=& -\Theta \left( { iM + \frac{1}{2}{\Lambda ^\dagger \Lambda}}
\right); 
 G  = -\Theta \Lambda^\dagger S;
\bar H  = \Lambda;
J  = S
\end{eqnarray}
and $S^\dagger S = I$. 
\end{definition} 
Here, the
notation $^\dagger$ represents complex conjugate transpose. In this
definition, if the system (\ref{sys0}) is physically realizable, then
the matrices $S$, $M$ and $\Lambda$ define a complex open harmonic
oscillator with scattering matrix $S$,   coupling operator $ L = \Lambda a $  and
a Hamiltonian operator $\mathcal{H}=a^\dagger M a$; e.g., see 
\cite{GZ00}, \cite{PAR92}, \cite{BHJ07},
\cite{JNP1} and \cite{EB05}. This definition is an extension of the definition given in
\cite{MaP3,JNP1} to allow for a general scattering matrix $S$; e.g.,
see \cite{GGY08,GJ09}.


The following theorem is an straightforward extension of Theorem 5.1 of \cite{MaP3} to
allow for a general scattering matrix $S$. 

\begin{theorem} (See \cite{MaP3}.)
\label{T1}
A complex linear quantum system of the form (\ref{sys0}) is physically
realizable if and only if there exists a matrix $\Theta=\Theta^\dagger
>0$ such that 
\begin{eqnarray}
\label{physrea}
F\Theta + \Theta F ^\dagger  +  G  G^\dagger  &=& 0;
 G  = -\Theta \bar H^\dagger J;
J^\dagger J  = I.
\end{eqnarray}
In this case, the corresponding Hamiltonian matrix \emph{M} is given by
\begin{equation}\label{Hamiltonian}
M = \frac{i}{2}\left( {\Theta^{-1}F  - F^\dagger\Theta^{-1}
} \right)
\end{equation}
and the corresponding coupling matrix $\Lambda$ is given by
\begin{equation}\label{coupling}\Lambda= \bar H. \end{equation}
\end{theorem}

\begin{definition}
\label{D2}
The linear complex quantum system (\ref{sys0}) is said to be \emph{lossless bounded real} if the following conditions hold:
\begin{enumerate}\item[i)] F is a Hurwitz matrix.
 \item[ii)]
$\Phi(s)=\bar H(sI-F)^{-1} G+J$ satisfies $\Phi(i\omega)^\dagger \Phi(i\omega)=I$ for all $ \omega \in \mathbb{R}.$ \end{enumerate}
\end{definition}

The following definition extends the standard linear systems notion of
minimal realization to linear complex quantum systems of the form
(\ref{sys0}); see also \cite{MaP3}.
\begin{definition}\label{D3}
A linear complex quantum system of the form (\ref{sys0}) is said to be \emph{minimal} if the
following conditions hold:
\begin{enumerate}\item[i)] {\em Controllability.} $a^\dagger F=
  \lambda a^\dagger$ for some $\lambda \in \mathbb{C}$ and $a^\dagger
   G =0$ implies $a=0$; \item[ii)] {\em Observability}. $Fa=\lambda
  a$ for some $\lambda \in \mathbb{C}$ and $\bar Ha=0$ implies
  $a=0$.
\end{enumerate}
\end{definition}

The following Theorem is an straightforward extension of Theorem 6.6 of \cite{MaP3} to
allow for a general scattering matrix $S$. 

\begin{theorem}\label{T2} A minimal linear complex quantum system of
  the form (\ref{sys0}) is physically realizable if and only if  the
  system is lossless bounded real.  
 \end{theorem}

\begin{definition}
\label{quantum_realization}
The complex linear  quantum system (\ref{sys0}) is said to be a {\em quantum
system realization} of a complex
transfer function matrix $K(s)$ if 
\begin{equation}
\label{tf0}
K(s) = \bar H(sI-F)^{-1} G + J.
\end{equation}
\end{definition}

 \section{The Cascade Cavity Realization Algorithm}
\label{sec:cascade_realization}
\begin{figure}[htbp]
\begin{center}
\psfrag{u=u0}{$\scriptscriptstyle u=u_0$}
\psfrag{u1=y0}{$\scriptscriptstyle u_1=y_0$}
\psfrag{u2=y1}{$\scriptscriptstyle u_2=y_1$}
\psfrag{un=ynm1}{$\scriptscriptstyle u_n=y_{n-1}$}
\psfrag{y=yn}{$\scriptscriptstyle y=y_n$}
\includegraphics[width=12cm]{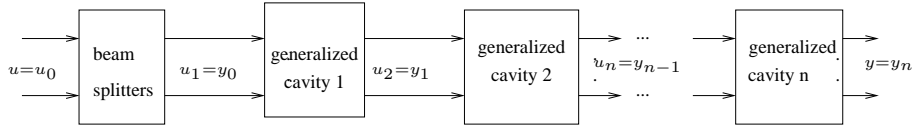}
\end{center}
\caption{Cascade  of $n$, generalized $m$
mirror cavities and a unitary operator.}
\label{F1}
\end{figure}

In this section, we recall the cascade cavity realization result of
\cite{PET08Aa,PET08A} and generalize it slightly to allow for quantum systems
with a more general scattering matrix. Indeed, given a linear quantum
system of the form (\ref{sys0}) with transfer function matrix (\ref{tf0}), we
can write $K(s) = J\tilde{K}(s)$ where
\begin{equation}
\label{tf}
\tilde{K}(s) =  H(sI-F)^{-1}G + I
\end{equation}
and $H = J^{-1} \bar H $. Corresponding to (\ref{tf})
is the linear quantum system 
\begin{eqnarray}\nonumber
\label{sys}
 da(t) &=& F a(t)dt + G du(t);   \nonumber \\
 dy(t) &=&  H a(t)dt +  du(t).
\end{eqnarray}
In order to obtain a physical realization of 
(\ref{tf0}), the result of \cite{PET08Aa,PET08A} can be applied to
transfer function matrix (\ref{tf}). Then a collection of
beamsplitters can be used to implement the unitary matrix $J$; e.g.,
see \cite{RZBB94}. This leads to a physical realization of the
transfer function matrix (\ref{tf}) as shown in Figure \ref{F1}.

An optical ring cavity consists of a number of partially
reflecting mirrors arranged to produce a traveling light wave when
coupled to a coherent light source; e.g., see \cite{BR04,GZ00}. If we
augment such a cavity by introducing phase-shifters
on the input and output channels, such a
cavity with $m$ mirrors, can be described by
a linear quantum system of the form
(\ref{sys}) as follows; see \cite{PET08Aa,PET08A}:
\begin{eqnarray}
\label{gen_cavity1}
da &=& pa dt - 
h^\dagger du; 
dy = ha dt + du
\end{eqnarray}
where
\begin{equation}
\label{realp}
p+p^* = -\gamma = -\sum_{i=1}^m \kappa_i = -h^\dagger h.
\end{equation}
Here $p = -\gamma/2+i\Delta$,
\begin{eqnarray*}
h &=& \left[\begin{array}{l}
h_1\\
h_2\\
\vdots \\
h_m
\end{array}\right] 
= \left[\begin{array}{l}
\sqrt{\kappa_1} e^{i\theta_1}\\
\sqrt{\kappa_2} e^{i\theta_2}\\
\vdots \\
\sqrt{\kappa_m} e^{i\theta_m}
\end{array}\right], 
du = \left[\begin{array}{l}
du_1\\
du_2\\
\vdots \\
du_m
\end{array}\right], ~ dy=\left[\begin{array}{l}
dy_1\\
dy_2\\
\vdots \\
dy_m
\end{array}\right].
\end{eqnarray*}
Furthermore,  any first order complex linear
quantum system of the form (\ref{gen_cavity1}), with non-zero $h \in
\mathbb{C}^{m}$ and satisfying (\ref{realp}), can be physically
realized as a generalized $m$ mirror cavity. In this case, the mirror
coupling coefficients and phase shifts are determined using a polar
coordinates description of the  elements of $h$. Also, the detuning
parameter $\Delta$ is determined from the imaginary part of the system
pole $p$. 

\begin{figure}[htbp]
\begin{center}
\psfrag{u=u1}{$\scriptstyle u=u_1$}
\psfrag{u2=y1}{$\scriptstyle u_2=y_1$}
\psfrag{un=ynm1}{$\scriptstyle u_n=y_{n-1}$}
\psfrag{y=yn}{$\scriptstyle y=y_n$}
\includegraphics[width=12cm]{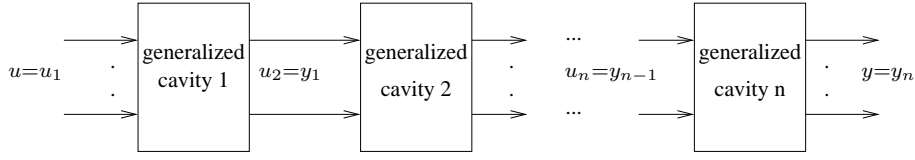}
\end{center}
\caption{Cascade  of $n$, generalized $m$
mirror cavities.}
\label{F2}
\end{figure}

The cascade cavity realization introduced in \cite{PET08Aa,PET08A} involves  a cascade interconnection of $n$, generalized $m$
mirror cavities as shown in Figure \ref{F2}. 
In this cascade system, the $i$th cavity is described by the following
QSDEs of the form
(\ref{sys}), (\ref{gen_cavity1}):
\begin{eqnarray}
\label{gen_cavity_i}
da_i &=& p_ia_i dt - 
H_i^\dagger du; 
dy = H_ia_i dt + du
\end{eqnarray}
where
\begin{equation}
\label{realp_i}
p_i+p_i^*  = -H_i^\dagger H_i.
\end{equation}
The cascade system
 is then described by 
a complex linear quantum
system of the form (\ref{sys}) where
\begin{eqnarray}
\label{cascade_matrices}
F&=&  \left[\begin{array}{cccc}
p_1 & 0  & \ldots  &  0 \\
-H_2^\dagger H_1 & p_2 & &   \\
\vdots &      & \ddots  &  \vdots \\
 &    & 
 & 0\\
-H_n^\dagger H_1 &  \ldots &  -H_n^\dagger H_{n-1} &
p_n \end{array}\right], 
G= -\left[\begin{array}{r}
H_1^\dagger \\
H_2^\dagger \\
\vdots \\
H_n^\dagger \\
\end{array}\right], \nonumber \\
H&=&\left[\begin{array}{rrrr}
H_1 & H_2 & \ldots & H_n
\end{array}\right], 
J= I.
\end{eqnarray}

Reference \cite{PET08Aa,PET08A} presents an algorithm for realizing a
physically realizable quantum system (\ref{sys}) with transfer
function (\ref{tf}) 
via a cascade of generalized cavities. We restrict attention to
quantum systems in which the transfer function (\ref{tf}) corresponds
to a minimal system (\ref{sys}) such that the eigenvalues of the
matrix $F$ are all distinct. In this case, it follows via a (complex
version of a) standard result from linear systems theory, that the system
(\ref{sys}) can be transformed into Modal Canonical Form;
e.g., see \cite{KAI80}. The complex linear quantum system in modal
canonical form is assumed to be as follows:
\begin{eqnarray}
\label{modal}
 d\tilde a(t) &=& \tilde F \tilde a(t)dt + \tilde G du(t);
 \nonumber \\
 dy(t) &=& \tilde H \tilde a(t)dt +  du(t)
\end{eqnarray}
where 
\begin{eqnarray}
\label{modal_matrices}
\tilde F &=& \left[\begin{array}{llll}
p_1 & 0 & \ldots & 0 \\
0 & p_2 &  & \vdots \\
\vdots & & \ddots & 0 \\
0 & \ldots & 0 & p_n
\end{array}\right];
\tilde G =  \left[\begin{array}{l} 
\tilde G_1 \\
\tilde G_2 \\
\vdots \\
\tilde G_n \\
\end{array}\right];
\tilde H = \left[\begin{array}{llll}
\tilde H_1 & \tilde H_2 & \vdots & \tilde H_n
\end{array}\right].
\end{eqnarray}
Also, it is assumed that in this realization,  the eigenvalues are
ordered so that $|p_1| \leq |p_2| \leq \ldots \leq |p_n|$.
Then, $K(s)$ satisfies the equation
\begin{equation}
\label{modal_tf}
K(s) = \tilde H(sI-\tilde F)^{-1}\tilde G + I.
\end{equation}

The algorithm proposed in \cite{PET08Aa,PET08A} is as follows:

\begin{description}
\item[Step 1:]
~Begin with a minimal modal canonical form realization (\ref{modal}),
(\ref{modal_matrices}) of 
the lossless bounded real transfer function matrix $K(s)$.
\item[Step 2:]
~Let
\begin{eqnarray}
\label{Hn}
\bar H_n &=& \tilde H_n, 
\alpha_n = -\frac{\bar H_n^\dagger \bar H_n}{p_n+p_n^*}, \nonumber \\
H_n &=& \frac{\bar H_n}{\sqrt{\alpha_n}},~t(n,n) =
\frac{1}{\sqrt{\alpha_n}}. 
\end{eqnarray}
\item[Step 3:]
~Calculate the quantities $H_n,H_{n-1},\ldots,H_1$,
$\alpha_n,\alpha_{n-1},\ldots,\alpha_1$, $t(i,j)$, for
$j=n,n-1,\ldots,1$ and $j \geq i$. These are calculated using
the following recursive formulas starting with the values determined
in Step 2 for $i=n$: 
\end{description}
\begin{eqnarray}
\label{Hbi}
\bar H_i &=& \left[I+\sum_{j=i+1}^n\frac{\tilde
    H_j}{p_j-p_i}\sum_{k=i+1}^j t(j,k)H_k^\dagger\right]^{-1}\tilde
H_i; \nonumber \\
\\
\label{alphai}
\alpha_i &=& -\frac{\bar H_i^\dagger \bar H_i}{p_i+p_i^*},~ H_i =
\frac{\bar H_i}{\sqrt{\alpha_i}}, \\
\label{tki}
t(k,i) &=& \frac{1}{p_i-p_k}\sum_{j=i+1}^k t(k,j)H_j^\dagger H_i \mbox{
  for } k=i+1,\ldots,n, \nonumber \\
\\
\label{tii}
t(i,i) &=& \frac{1}{\sqrt{\alpha_i}}.
\end{eqnarray} 
\begin{description}
\item[Step 4:]
~Set $t(k,i) = 0$ for $k < i$ and define an $n\times n$ transformation
matrix $T$ whose $(i,j)$th element is $t(i,j)$. 
\end{description}

The following theorem is presented in \cite{PET08Aa,PET08A}. 

\begin{theorem}
\label{T3}
Consider an $m \times m$ lossless bounded real complex transfer
function matrix $K(s)$ with a minimal modal canonical form quantum
realization (\ref{modal}), (\ref{modal_matrices}) such that the
eigenvalues of the matrix $\tilde F$ are all distinct and that all of
the matrix inverses exist in equation (\ref{Hbi}) when the above
algorithm is applied to the 
system (\ref{modal}), (\ref{modal_matrices}). 
Then the vectors
$H_1,~H_2,\ldots,H_n$ defined in the above algorithm together with the
eigenvalues $p_1,~p_2,\ldots,p_n$ define an equivalent cascade quantum
realization (\ref{sys}), (\ref{cascade_matrices}) for the transfer
function matrix $K(s)$. Furthermore, this system is such that the
condition (\ref{realp_i}) is 
satisfied for all $i$. Moreover, the matrices $\{F,G,H,I\}$
defining this cascade quantum realization are related to the matrices
$\{\tilde F, \tilde G, \tilde H, I\}$ defining the modal quantum
realization (\ref{modal}), (\ref{modal_matrices}) according to the formulas:
\begin{equation}
\label{ss_equiv}
\tilde F = TFT^{-1},~\tilde G = T G,~\tilde H = HT^{-1}
\end{equation}
where the matrix $T$ is defined in the above algorithm. 
\end{theorem}

The physical realization of   (\ref{tf0}) corresponds to writing
\begin{equation}
\label{cascade_realization}
K(s) = J\tilde{K}_n(s)\tilde{K}_{n-1}(s)\ldots\tilde{K}_1(s)
\end{equation}
where each transfer function matrix $\tilde{K}_i(s)$ is a first order
transfer function matrix corresponding to an optical cavity described by
a QSDE of the form (\ref{gen_cavity_i}).

\section{The Main Result}
\label{sec:main_result}
Our proposed method for obtaining an approximate model for a complex
linear quantum system (\ref{sys0}) with transfer function matrix
(\ref{tf0}) involves truncating the cascade realization
(\ref{cascade_realization}) to obtain the approximate transfer function matrix
\begin{equation}
\label{cascade_approximation}
K_a(s) = J_a\tilde{K}_r(s)\tilde{K}_{r-1}(s)\ldots\tilde{K}_1(s)
\end{equation}
where 
$
J_a = J\tilde{K}_{n}(0)\tilde{K}_{n-1}(0)\ldots\tilde{K}_{r+1}(0)
$
and
$r < n$ is the order of the approximate model. It follows from this
construction that  $K_a(s)$ is lossless
bounded real and hence physically realizable. Indeed, since the
transfer function matrix
$J\tilde{K}_{n}(s)\tilde{K}_{n-1}(s)\ldots\tilde{K}_{r+1}(s)$ is
lossless bounded real, it follows that the matrix $J_a$ will be
unitary. Therefore, $K_a(s)$ will be lossless bounded real since
$\tilde{K}_r(s)\tilde{K}_{r-1}(s)\ldots\tilde{K}_1(s)$ is lossless bounded
real. 

In order to construct a state space realization of the reduced
dimension transfer function matrix $K_a(s)$, note that it follows from
the development in Section \ref{sec:cascade_realization} that the
transfer function matrix $\tilde K_a(s) =
\tilde{K}_r(s)\tilde{K}_{r-1}(s)\ldots\tilde{K}_1(s)$ has a state space
realization of the form (\ref{sys}) defined by the matrices
\begin{eqnarray}
\label{cascade_matrices1}
\tilde F_a&=&  \left[\begin{array}{cccc}
p_1 & 0  & \ldots  &  0 \\
-H_2^\dagger H_1 & p_2 & &   \\
\vdots &      & \ddots  &  \vdots \\
 &    & 
 & 0\\
-H_r^\dagger H_1 &  \ldots &  -H_r^\dagger H_{r-1} &
p_r \end{array}\right], 
\tilde G_a= -\left[\begin{array}{r}
H_1^\dagger \\
H_2^\dagger \\
\vdots \\
H_r^\dagger \\
\end{array}\right], \nonumber \\
\tilde H_a&=&\left[\begin{array}{rrrr}
H_1 & H_2 & \ldots & H_r
\end{array}\right].
\end{eqnarray}
Also, the transfer function matrix $K_b(s) =
\tilde{K}_{n}(s)\tilde{K}_{n-1}(s)\ldots\tilde{K}_{r+1}(s)$ has a 
state space
realization of the form (\ref{sys}) defined by the matrices
\begin{eqnarray}
\label{cascade_matrices2}
F_b&=&  \left[\begin{array}{cccc}
p_{r+1} & 0  & \ldots  &  0 \\
-H_{r+2}^\dagger H_{r+1} & p_{r+2} & &   \\
\vdots &      & \ddots  &  \vdots \\
 &    & 
 & 0\\
-H_n^\dagger H_{r+1} &  \ldots &  -H_n^\dagger H_{n-1} &
p_n \end{array}\right], 
G_b= -\left[\begin{array}{r}
H_{r+1}^\dagger \\
H_{r+2}^\dagger \\
\vdots \\
H_n^\dagger \\
\end{array}\right], \nonumber \\
H_b&=&\left[\begin{array}{rrrr}
H_{r+1} & H_{r+2} & \ldots & H_n
\end{array}\right].
\end{eqnarray}
Hence, the matrix $J_a$ is given by 
$
J_a = J\left(I-H_bF_b^{-1}G_b\right)
$
and the reduced dimension transfer function matrix $K_a(s)$ has a
state space realization of the form (\ref{sys0}) defined by the
matrices $\left(\tilde F_a,\tilde G_a,J_a\tilde H_a,J_a\right)$. 

The ordering of
the eigenvalues in the cascade realization (\ref{modal_matrices})
means that this model is expected to be a good approximation of
original model at low frequencies $\omega << |p_{r+1}|.$ The
corresponding error system is defined by
\begin{eqnarray}
\label{approximation_error}
K_e(s) &=& K(s) - K_a(s) \nonumber \\
&=&
J\left(\begin{array}{l}\tilde{K}_{n}(s)\tilde{K}_{n-1}(s)\ldots\tilde{K}_{r+1}(s) \\
  -
  \tilde{K}_{n}(0)\tilde{K}_{n-1}(0)\ldots\tilde{K}_{r+1}(0)\end{array}\right)
\tilde{K}_r(s)\tilde{K}_{r-1}(s)\ldots\tilde{K}_1(s).
\end{eqnarray}

We now present
a result which bounds the induced matrix norm of the approximation error $\|K_e(j\omega)\|$ as
a function of frequency.  This bound will be defined in terms of the
following quantities:
\begin{eqnarray*}
B_1(\omega) &=& \omega \sum_{i=r+1}^n
\frac{-(p_i+p_i^*)}{|p_i|.|p_i-j\omega|}; \nonumber \\
B_2(\omega) &=& \omega \sum_{i_1=r+1}^n\sum_{i_2=i_1+1}^n
C_{i_1,i_2}(\omega);\nonumber \\
\nonumber \\
& \vdots & \nonumber \\
B_k(\omega) &=&
\end{eqnarray*}
\[
\omega \sum_{i_1=r+1}^n\sum_{i_2=i_1+1}^n \ldots \sum_{i_k=i_{k-1}+1}^n
C_{i_1,i_2,\ldots,i_k}(\omega);
\]
\begin{eqnarray} 
\label{bounds1}
& \vdots & \nonumber \\
B_{n-r}(\omega) &=& 
\omega C_{r+1,r+2,\ldots,n}(\omega) \hspace{1.5cm}
\end{eqnarray}
where
\[
\small
C_{i_1,i_2,\ldots,i_k}(\omega) = 
\left|\begin{array}{l}
\scriptscriptstyle
(-j\omega)^{k-1} + (-j\omega)^{k-2}\sum_{m=1}^kp_{i_m}+ \\
\scriptscriptstyle
(-j\omega)^{k-3}\sum_{m_1=1}^k\sum_{m_2=m_1+1}^kp_{i_{m_1}}p_{i_{m_2}}+\\
\scriptscriptstyle
(-j\omega)^{k-4}\sum_{m_1=1}^k\sum_{m_2=m_1+1}^k\sum_{m_3=m_2+1}^k
p_{i_{m_1}}p_{i_{m_2}}p_{i_{m_3}}\\
\scriptscriptstyle
+ \ldots + \\
\scriptscriptstyle
(-j\omega)^{k-p-1}\sum_{m_1=1}^k\sum_{m_2=m_1+1}^k\ldots \sum_{m_p=m_{p-1}+1}^k
\prod_{q=1}^pp_{i_{m_q}}\\
\scriptscriptstyle
+ \ldots + \\
\scriptscriptstyle
(-j\omega)^{k-p-1}\sum_{m_1=1}^k\sum_{m_2=m_1+1}^k\ldots \sum_{m_{k-1}=m_{k-2}+1}^k
\prod_{q=1}^{k-1}p_{i_{m_q}}
\end{array}\right| 
\scriptstyle 
\frac{\prod_{l=1}^k-(p_{i_l}+p_{i_l}^*)}{\prod_{l=1}^k|p_{i_l}|.|p_{i_l}
  - j\omega |}.\hspace{5.5cm}
\]
\begin{equation}
\label{Cis}
\end{equation}

\begin{theorem}
\label{T5}
Consider a physically realizable linear complex quantum system of the
form (\ref{sys0}) and corresponding transfer function matrix
(\ref{tf0}). Suppose this system has a cascade cavity realization
(\ref{cascade_realization}) and a corresponding approximate transfer function
matrix $K_a(s)$ defined in (\ref{cascade_approximation}). Then
$K_a(s)$ is physically realizable and the corresponding approximation
error transfer function matrix $K_e(s)$ defined in
(\ref{approximation_error}) satisfies the bound
\begin{equation}
\label{bound1}
\|K_e(j\omega)\| \leq \sum_{k=1}^{n-r}B_k(\omega)
\end{equation}
for all $\omega \geq 0$ where the quantities $B_k(\omega)$ are defined
in (\ref{bounds1}), (\ref{Cis}).
\end{theorem}

{\em Proof.}
The fact that $K_a(s)$ is physically realizable follows from its
definition as discussed above. Now for any $\omega \geq 0$, it follows
from (\ref{approximation_error}) that
\begin{eqnarray}
\label{error_bound1}
\|K_e(j\omega)\| &\leq& \|J\| 
\left\|\begin{array}{l}\tilde{K}_{n}(j\omega)\tilde{K}_{n-1}(j\omega)\ldots\tilde{K}_{r+1}(j\omega)\\
  -
  \tilde{K}_{n}(0)\tilde{K}_{n-1}(0)\ldots\tilde{K}_{r+1}(0)\end{array}\right\|\nonumber
\\
&& \times
\|\tilde{K}_r(j\omega)\tilde{K}_{r-1}(j\omega)\ldots\tilde{K}_1(j\omega)\|\nonumber\\
&=& \left\|\begin{array}{l}\tilde{K}_{n}(j\omega)\tilde{K}_{n-1}(j\omega)\ldots\tilde{K}_{r+1}(j\omega)
  -
  \tilde{K}_{n}(0)\tilde{K}_{n-1}(0)\ldots\tilde{K}_{r+1}(0)\end{array}\right\|
\nonumber \\
\end{eqnarray}
using the fact that the transfer function matrix
$\tilde{K}_r(s)\tilde{K}_{r-1}(s)\ldots\tilde{K}_1(s)$ is lossless
bounded real and the matrix $J$ is unitary. 

We now consider the transfer function matrices
\begin{eqnarray*}
\tilde K_e(s) &=& \tilde{K}_{n}(s)\tilde{K}_{n-1}(s)\ldots\tilde{K}_{r+1}(s)
-  \tilde{K}_{n}(0)\tilde{K}_{n-1}(0)\ldots\tilde{K}_{r+1}(0)
\end{eqnarray*}
and 
$
\hat K_e(s) = \tilde{K}_{n}(s)\tilde{K}_{n-1}(s)\ldots\tilde{K}_{r+1}(s).
$
Also, note that it follows from (\ref{gen_cavity_i}) that each transfer function
matrix $\tilde{K}_{i}(s)$ is of 
the form
$
\tilde{K}_{i}(s) = I + \frac{H_iH_i^\dagger}{p_i-s}.
$
Hence, we can write
$
\hat K_e(s) = \prod_{i=n}^{r+1}\left(I + \frac{H_iH_i^\dagger}{p_i-s}\right).
$
From this it follows that we can write
$
\hat K_e(s) = I + \sum_{k=1}^{n-r}T_k(s)
$
and 
\begin{equation}
\label{Ke_tilde}
\tilde K_e(s) = \sum_{k=1}^{n-r}\left(T_k(s)-T_k(0)\right)
\end{equation}
where the transfer function matrices $T_{1}(s), T_{2}(s), \ldots,
T_{n-r}(s)$ are defined as follows:
\begin{eqnarray*}
T_{1}(s) &=& \sum_{i=r+1}^n \frac{H_i H_i^\dagger}{p_i - s};\\
T_{2}(s) &=& \sum_{i_1=r+1}^n\sum_{i_2=i_1+1}^n 
\frac{H_{i_2} H_{i_2}^\dagger H_{i_1} H_{i_1}^\dagger }
{\left(p_{i_1} - s\right)\left(p_{i_2} - s\right)}; \\
&\vdots& \\
T_{k}(s) &=& \sum_{i_1=r+1}^n\sum_{i_2=i_1+1}^n \ldots \sum_{i_k=i_{k-1}+1}^n
\prod_{l=k}^{1}
\frac{H_{i_l} H_{i_l}^\dagger}
{p_{i_l} - s}; \\
&\vdots& \\
T_{n-r}(s) &=& \prod_{i=n}^{r+1}\frac{H_{i} H_{i}^\dagger}{p_{i} - s}. 
\end{eqnarray*}
Now it follows from (\ref{Ke_tilde}) and the triangle inequality that
\begin{equation}
\label{tildeKebound}
\|\tilde K_e(j\omega)\| \leq \sum_{k=1}^{n-r}\|T_k(j\omega)-T_k(0)\|
\end{equation}
for all $\omega \geq 0$. We now consider each of the terms
$\tilde T_k(j\omega) =  T_k(j\omega)-T_k(0).$
Indeed, for any $\omega \geq 0$, we obtain
\begin{eqnarray*}
\tilde T_{1}(j\omega) &=& \sum_{i=r+1}^n 
\frac{j\omega H_iH_i^\dagger}{p_i\left(p_i - j\omega\right)}\nonumber
\\
\tilde T_{2}(j\omega) &=& \sum_{i_1=r+1}^n\sum_{i_2=i_1+1}^n 
H_{i_2} H_{i_2}^\dagger H_{i_1} H_{i_1}^\dagger 
 j\omega  S_{i_1,i_2}(j\omega) \nonumber \\
&\vdots& \nonumber\\
\tilde T_{k}(j\omega) &=& 
\end{eqnarray*}
\[
\sum_{i_1=r+1}^n\sum_{i_2=i_1+1}^n \ldots \sum_{i_k=i_{k-1}+1}^n
\prod_{l=k}^{1}H_{i_l} H_{i_l}^\dagger j\omega
S_{i_1,\ldots,i_k}(j\omega);
\]
\begin{eqnarray}
\label{tildeTs}
&\vdots& \nonumber\\
\tilde T_{n-r}(j\omega) &=& \prod_{i=n}^{r+1}H_{i} H_{i}^\dagger
j\omega S_{r+1,\ldots,n}(j\omega) \hspace{2.5cm}
\end{eqnarray}
where
\begin{eqnarray*}
\small
\lefteqn{S_{i_1,i_2,\ldots,i_k}(j\omega)} \\
&& = 
\left[\begin{array}{l}
\scriptscriptstyle
(-j\omega)^{k-1} + (-j\omega)^{k-2}\sum_{m=1}^kp_{i_m}+ \\
\scriptscriptstyle
(-j\omega)^{k-3}\sum_{m_1=1}^k\sum_{m_2=m_1+1}^kp_{i_{m_1}}p_{i_{m_2}}+\\
\scriptscriptstyle
(-j\omega)^{k-4}\sum_{m_1=1}^k\sum_{m_2=m_1+1}^k\sum_{m_3=m_2+1}^k
p_{i_{m_1}}p_{i_{m_2}}p_{i_{m_3}}\\
\scriptscriptstyle
+ \ldots + \\
\scriptscriptstyle
(-j\omega)^{k-p-1}\sum_{m_1=1}^k\sum_{m_2=m_1+1}^k\ldots \sum_{m_p=m_{p-1}+1}^k
\prod_{q=1}^pp_{i_{m_q}}\\
\scriptscriptstyle
+ \ldots + \\
\scriptscriptstyle
(-j\omega)^{k-p-1}\sum_{m_1=1}^k\sum_{m_2=m_1+1}^k\ldots \sum_{m_{k-1}=m_{k-2}+1}^k
\prod_{q=1}^{k-1}p_{i_{m_q}}
\end{array}\right]
\scriptstyle
\frac{1}{\prod_{l=1}^kp_{i_l}(p_{i_l}
  - j\omega)}. \hspace{6cm} 
\end{eqnarray*}
\begin{equation}
\label{tildeCis}
\end{equation}
Now it follows from (\ref{realp_i}) that 
$
\|H_{i} H_{i}^\dagger\| = H_{i}^\dagger H_{i} = -p_i-p_i^*
$
for $i = r+1,r+2,\ldots,n$. Hence, applying the triangle inequality
and the Cauchy-Schwartz inequality to (\ref{tildeTs}), it follows that
for any $\omega \geq 0$
\begin{eqnarray}
\label{tildeTs_bounds}
\|\tilde T_{1}(j\omega)\| &\leq & B_1(\omega);\nonumber
\\
\|\tilde T_{2}(j\omega)\| &\leq & B_2(\omega);\nonumber \\
&\vdots& \nonumber\\
\|\tilde T_{k}(j\omega)\| &\leq & B_k(\omega);\nonumber\\
&\vdots& \nonumber\\
\|\tilde T_{n-r}(j\omega)\| &\leq & B_{n-r}(\omega).
\end{eqnarray}
These bounds combined with (\ref{tildeKebound}) and (\ref{error_bound1}) lead to the inequality (\ref{bound1}). 
\hfill $\Box$

\begin{remark}
The quantity $\sum_{k=1}^{n-r}B_k(\omega)$ in bound (\ref{bound1}) is probably too complicated to calculate for all cases except when $n-r$ is equal to one or two. However, we can obtain some good approximations to this quantity which apply at low frequencies. Indeed, for $\omega << |p_{r+1}|$, we obtain 
\[
\sum_{k=1}^{n-r}B_k(\omega) \approx B_1(\omega)= \omega \sum_{i=r+1}^n
\frac{-(p_i+p_i^*)}{|p_i|.|p_i-j\omega|}
\]
for all $\omega \geq 0$. 
Furthermore since 
$
\sup_{\omega \geq 0} |p_i-j\omega|= -\frac{1}{2}(p_i+p_i^*),
$
we obtain the following useful upper bound on $ B_1(\omega)$:
$
B_1(\omega) \leq 2 \omega \sum_{i=r+1}^n
\frac{1}{|p_i|}
$
for all $\omega \geq 0$. 
\end{remark}

\section{Illustrative Example}
\label{sec:example}
\begin{figure}[htbp]
\begin{center}
\includegraphics[width=9.5cm]{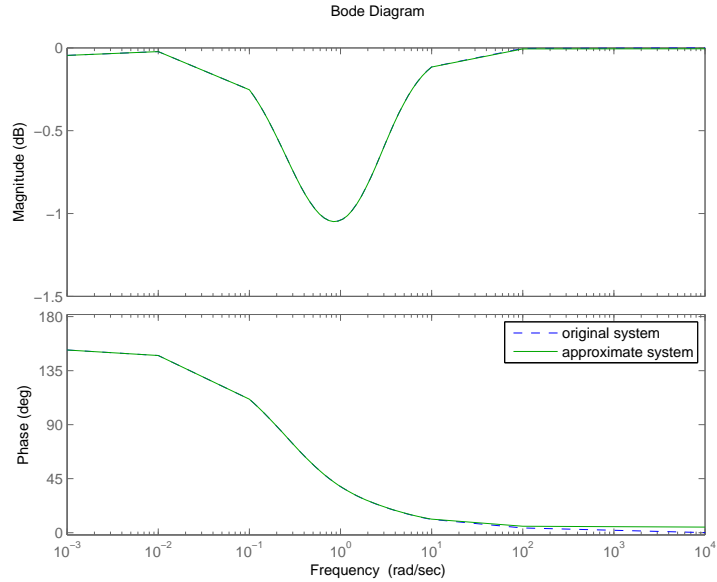}
\end{center}
\caption{Bode plot of original and approximate system from input 1 to output 1.}
\label{F3}
\end{figure}

\begin{figure}[htbp]
\begin{center}
\includegraphics[width=9.5cm]{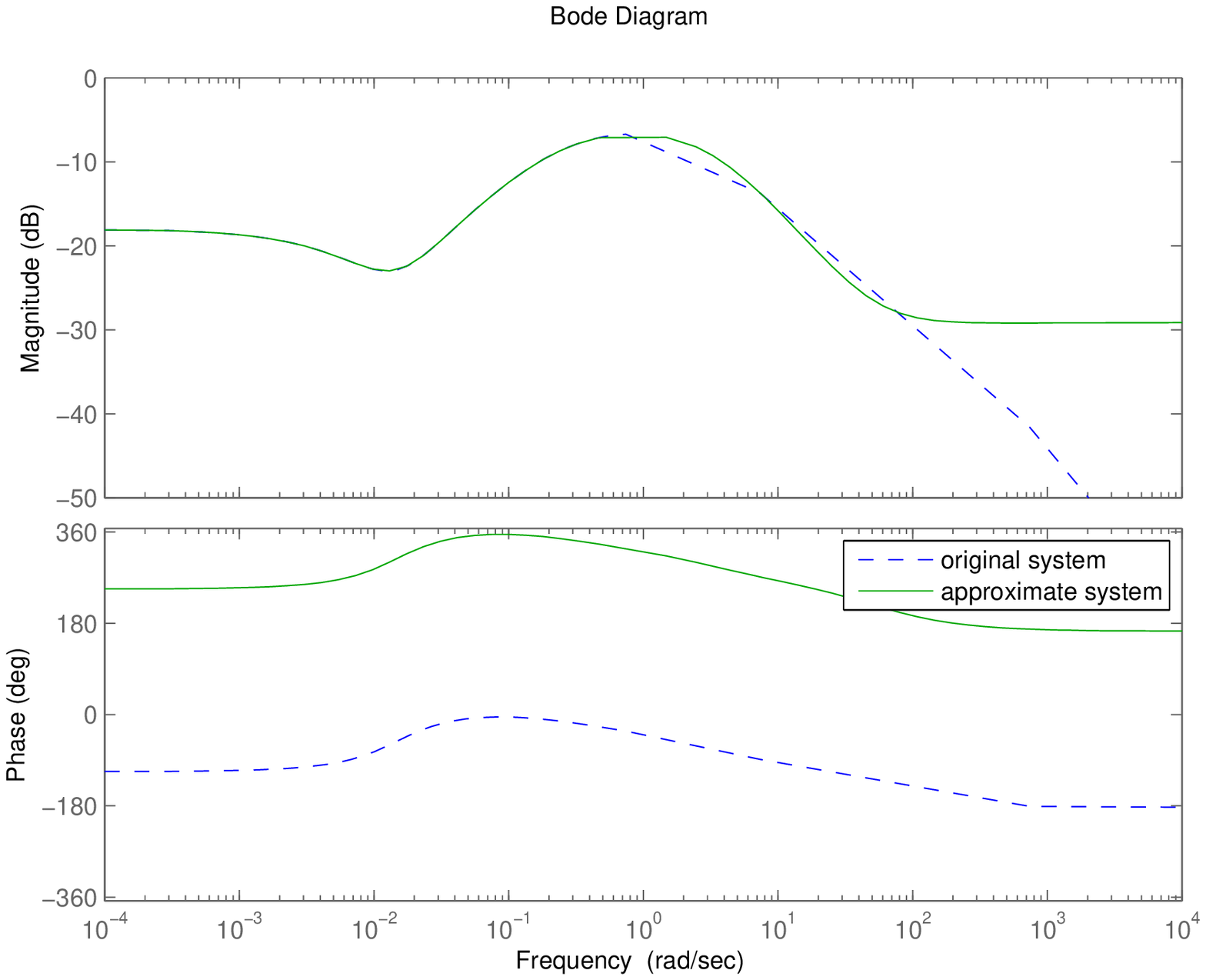}
\end{center}
\caption{Bode plot of original and approximate system from input 2 to output 1.}
\label{F4}
\end{figure}

To obtain our initial physically realizable quantum system, we start
with a random Hamiltonian matrix $M > 0$, a random coupling matrix $\Lambda$, and a random
commutation matrix $\Theta > 0$
defined as follows:
\begin{eqnarray*}
\small
M = \hspace{12cm} \\
%
\left[\begin{array}{lll}
   3.3314             &2.5448 + 0.8204j   &2.4007 + 1.1592j    \\
   2.5448 - 0.8204j   &3.3994             &2.7136 + 0.4185j    \\
   2.4007 - 1.1592j   &2.7136 - 0.4185j   &4.1258              \\
   3.6470 - 1.3066j   &3.7009 + 0.1246j   &4.2612 + 0.3611j    \\
   1.9949 - 1.8970j   &2.6090 - 1.4039j   &3.5647 - 1.0224j       
\end{array}\right.  \\
\left.\begin{array}{ll}
3.6470 + 1.3066j   &1.9949 + 1.8970j\\
3.7009 - 0.1246j   &2.6090 + 1.4039j\\
4.2612 - 0.3611j   &3.5647 + 1.0224j\\
5.9568             &4.1173 + 1.3450j\\
4.1173 - 1.3450j   &3.9970          
\end{array}\right],
\end{eqnarray*}
\begin{eqnarray*}
\small
\Theta =\hspace{12cm}
\end{eqnarray*}
\begin{eqnarray*}
\small
\left[\begin{array}{lll}
   1.8356             &2.3408 - 0.3287j   &1.8732 + 0.0757j   \\
   2.3408 + 0.3287j   &3.9779             &3.3007 + 0.4982j   \\
   1.8732 - 0.0757j   &3.3007 - 0.4982j   &3.8582             \\
   1.8750 - 0.0488j   &2.8626 - 0.2561j   &2.5536 + 0.7201j   \\
   1.5306 - 0.3665j   &2.3995 - 1.1168j   &3.1242 - 0.8781j         
\end{array}\right. \\
\left.\begin{array}{ll}
1.8750 + 0.0488j   &1.5306 + 0.3665j \\
2.8626 + 0.2561j   &2.3995 + 1.1168j \\
2.5536 - 0.7201j   &3.1242 + 0.8781j \\
3.0974             &1.9480 + 0.9779j \\
1.9480 - 0.9779j   &3.0319          
\end{array}\right],
\end{eqnarray*}
\begin{eqnarray*}
\small
\Lambda =\hspace{12cm}
\end{eqnarray*}
\begin{eqnarray*}
\small
\left[\begin{array}{lll}
  -1.0106 + 0.0000j   &0.5077 + 1.0950j   &0.5913 + 0.4282j  \\
   0.6145 - 0.3179j   &1.6924 - 1.8740j  &-0.6436 + 0.8956j  
\end{array}\right. \\
\left.\begin{array}{ll}
 0.3803 + 0.7310j  &-0.0195 + 0.0403j \\
-1.0091 + 0.5779j  &-0.0482 + 0.6771j
\end{array}\right].
\end{eqnarray*}
Also, we chose the  scattering matrix $S=I$. This leads to a corresponding system of the form (\ref{sys0}) where the matrices are defined as in (\ref{harmonic}). The eigenvalues of the resulting matrix $F$ are $s=-0.0038 - 0.0181j$, $s=-0.2103 - 0.1040j$, $s=-0.1674 - 1.1066j$, $s=-3.0388 - 0.9275j$, $s=-10.8541-225.9473j$. Clearly, the last eigenvalue has a much larger absolute value than all of the others and so we will apply our algorithm to approximate this fifth order system by a fourth order system. Bode plots comparing the original system frequency response with the reduced dimension system frequency response are shown in Figures \ref{F1}-\ref{F4}. These Bode plots indicate that the proposed method gives a good approximation at low frequencies. Also, it follows from the construction that the reduced dimension system is lossless bounded real and so physically realizable. 

\begin{figure}[htbp]
\begin{center}
\includegraphics[width=9.5cm]{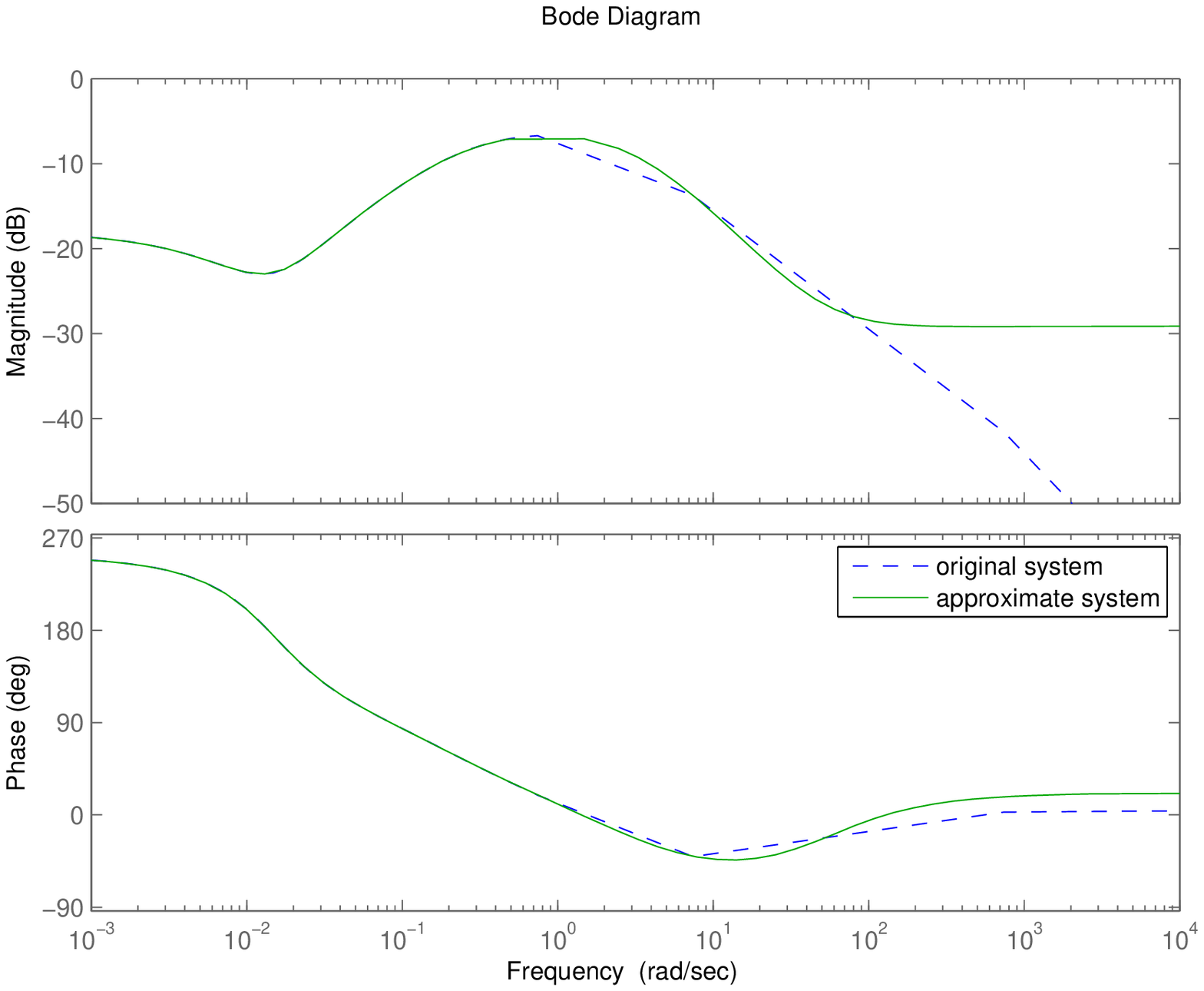}
\end{center}
\caption{Bode plot of original and approximate system from input 1 to output 2.}
\label{F5}
\end{figure}

\begin{figure}[htbp]
\begin{center}
\includegraphics[width=9.5cm]{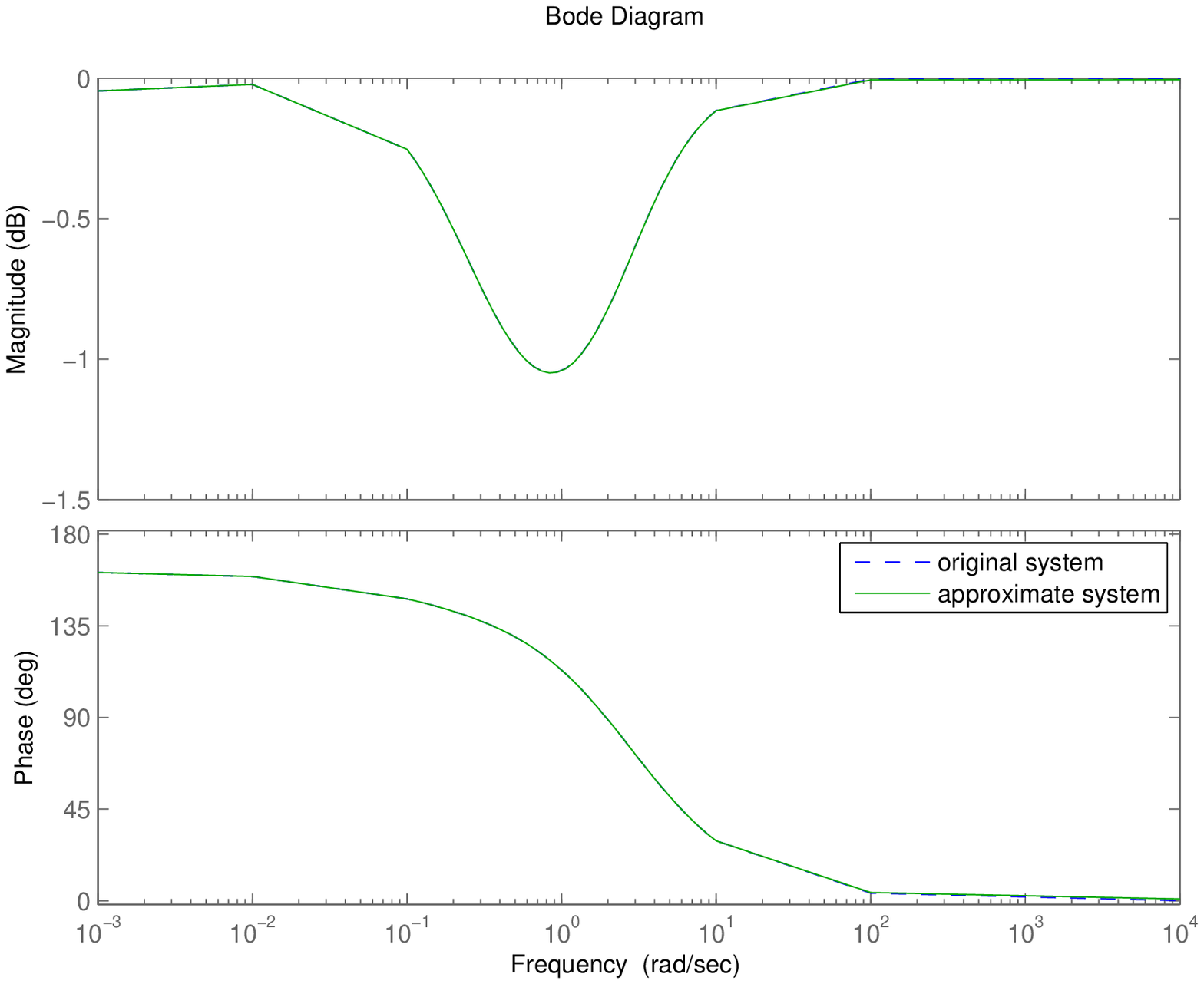}
\end{center}
\caption{Bode plot of original and approximate system from input 2 to output 2.}
\label{F6}
\end{figure}

\begin{figure}[htbp]
\begin{center}
\includegraphics[width=9.5cm]{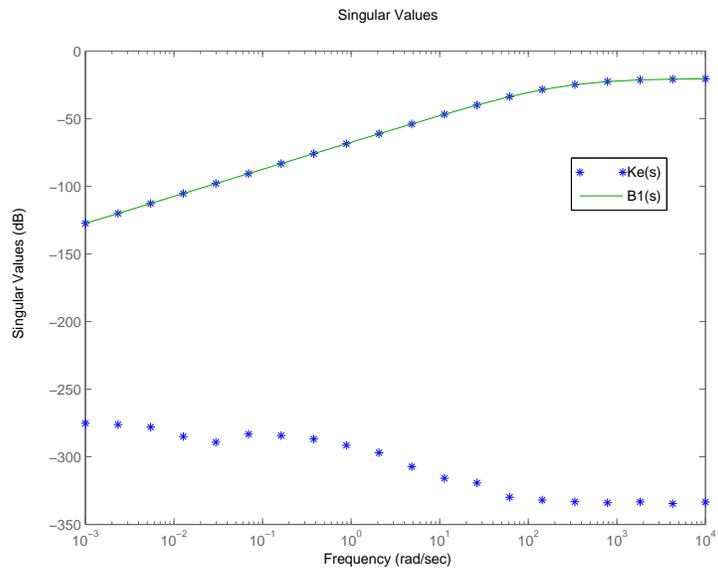}
\end{center}
\caption{Singular value plot of the error transfer function matrix $K_e(s)$ and the error bound $B_1(\omega)$.}
\label{F7}
\end{figure}

In Figure \ref{F7}, we show the singular value plot of the error transfer function matrix $K_e(s) = K(s) - K_a(s)$ along with the error bound defined by $B_1(\omega)$. In this example, we see that the error bound is in fact exact since we only reduced the dimension of the original system by one.  

\section{Conclusions}
In this paper, we have presented a method of approximating a class of linear complex quantum systems in such a way that the property of physical realizability (which is equivalent to the strict bounded real property in this case) is preserved. The paper presents a bound on the approximation error which shows that the approximation is accurate at low frequencies. 

\section{acknowledgment}
The author is wishes to acknowledge Professor Hideo Mabuchi who
suggested that the physical realization algorithm of \cite{PET08Aa,PET08A}
may also be useful in the approximation of linear quantum systems. 


\end{document}